\documentclass[11pt]{article}

\usepackage[margin=0.86in]{geometry}
\usepackage{amsmath,amssymb}
\usepackage{booktabs}
\usepackage{graphicx}
\usepackage{microtype}
\usepackage{tabularx}
\usepackage{longtable}
\usepackage{array}
\usepackage{xcolor}
\usepackage{caption}
\usepackage{subcaption}
\usepackage{tikz}
\usepackage{graphicx}
\usepackage{xcolor}
\usepackage{tikz}
\usepackage{amsmath}
\usepackage{float}
\usepackage{placeins}
\usepackage{listings}

\usetikzlibrary{positioning,arrows.meta,calc,decorations.pathreplacing}

\definecolor{depth0fill}{RGB}{166, 54, 46}    
\definecolor{depth1fill}{RGB}{205, 103, 55}   
\definecolor{depth2fill}{RGB}{224, 164, 72}   
\definecolor{depth3fill}{RGB}{235, 205, 135}  

\usetikzlibrary{arrows.meta,positioning,shapes.geometric,fit}
\usepackage[numbers,sort&compress]{natbib}
\usepackage[hidelinks]{hyperref}
\usepackage{url}
\hypersetup{
  pdftitle={Early Forecasting of Reply Chain Networks Propagation Growth for Misinformation Triage},
  pdfauthor={Ansh Gupta, Abhiram Gorle, Aayush Rajesh, and Tsachy Weissman},
  pdfsubject={Early forecasting of recorded misinformation propagation growth using structural and temporal entropy},
  pdfkeywords={misinformation, propagation forecasting, virality, structural entropy, temporal entropy, human-in-the-loop}
}

\definecolor{virality}{HTML}{0072B2}
\definecolor{veracity}{HTML}{D55E00}
\definecolor{triage}{HTML}{009E73}
\definecolor{softblue}{HTML}{E7F2F8}
\definecolor{softorange}{HTML}{FCEDE5}
\definecolor{softgreen}{HTML}{E8F5EF}

\newcommand{\R}{\mathbb{R}}

\newcolumntype{Y}{>{\raggedright\arraybackslash}X}
\title{Forecasting the Growth of Social Media Information Cascades: Towards Human-in-the-Loop Misinformation Triage}

\author{
Ansh Gupta$^{1}$ \and
Abhiram Gorle$^{2}$ \and
Aayush Rajesh$^{2}$ \and
Tsachy Weissman$^{2}$\\[0.45em]
\small $^{1}$The Overlake School \qquad $^{2}$Stanford University
}
\date{}

\begin{document}
\maketitle

\begin{abstract} 

Limited review teams must identify which emerging claims are likely to keep growing before their eventual reach is known. We center early misinformation triage on this continuation-forecasting problem: predicting subsequent recorded propagation-tree growth from the first 30 minutes of activity. On FibVID, we compare early node count with structural depth entropy, temporal arrival entropy, and their pair while keeping all propagation trees from each original claim in one partition. Across 352 test trees from 59 claim groups separate from training, the combined model improves every reported growth metric, raising $R^2$ for log-transformed future growth from 0.307 to 0.323 and reducing log-MAE by 2.4\% (95\% claim-bootstrap CI, $-0.4$\% to 5.2\%). The gain is especially pronounced among 97 high-activity trees: $R^2$ rises from 0.248 to 0.395, Spearman's $\rho$ from 0.394 to 0.529, and log-MAE falls by 11.7\% (95\% CI, $-1.9$\% to 23.9\%). Complementing the 30-minute growth forecast, we analyze the first 15 replies in 563 PHEME threads. In this eligible cohort, the 15th reply arrives after a median of 28.8 minutes; 52.0\% reach the fixed reply prefix within 30 minutes and 71.6\% within one hour, placing the response analysis on a comparable early-triage timescale. Even without the LLM-generated factual-accuracy dimension, the remaining stance, communicative, and affective state composition retains cross-event ranking signal (ROC--AUC 0.538); including that dimension in a secondary comparison increases ROC--AUC to 0.562. In a separate Check-COVID evaluation of 229 claims, reciprocal-rank fusion retrieves a gold evidence document within the top five for 74.2\% of claims and within the top 20 for 94.3\%; sentence reranking reaches Recall@20 of 58.1\%, showing strong document retrieval while identifying fine-grained evidence selection as the next opportunity. We propose an integrated human-review system that brings these early forecasts, response patterns, and retrieved evidence together for misinformation triage relying on the potential virality of claims.

\end{abstract}

\section{Introduction}

Online misinformation can acquire a large audience before a definitive fact-check is available. The central operational challenge is therefore to identify which claims are likely to continue growing while there is still time for a reviewer to act. Systems that wait for a mature cascade observe the outcome they are supposed to anticipate. We address this challenge by forecasting subsequent recorded propagation growth from an early, fixed observation window, while keeping social-response and evidence signals available as complementary reviewer context.

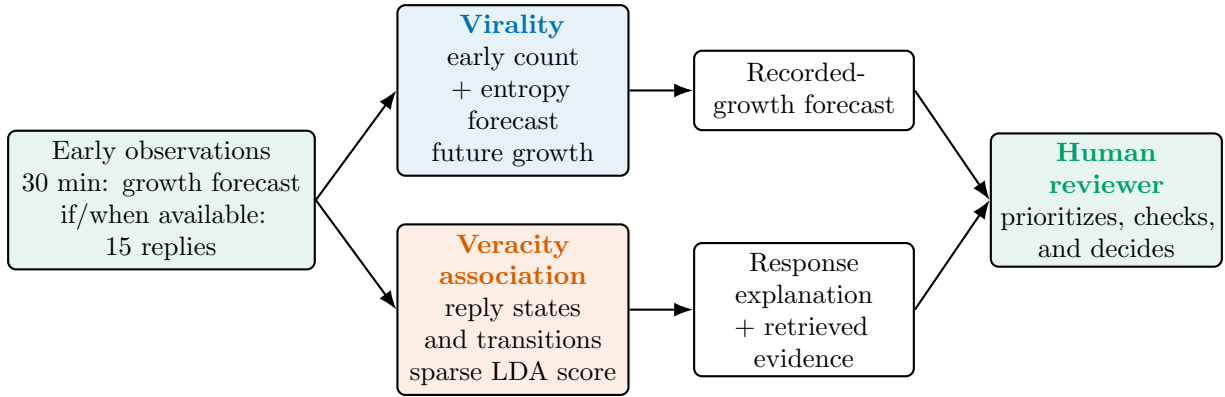
\begin{figure}[t]
\centering
\begin{tikzpicture}[
    font=\small,
    box/.style={
        rounded corners=3pt,
        draw,
        thick,
        align=center,
        minimum height=1.05cm,
        text width=3.0cm
    },
    arr/.style={-{Latex[length=2.5mm]},thick}
]


\node[box,fill=softgreen,text width=3.8cm] 
    (prefix) at (0,0)
    {Early observations\\
     \mbox{30 min: growth forecast}\\
     \mbox{if/when available:}\\
     15 replies};

\node[box,fill=softblue,text width=2.8cm] 
    (vir) at (4.65,1.45)
    {\textcolor{virality}{\textbf{Virality}}\\
     early count + entropy\\
     forecast\\future growth};

\node[box,fill=softorange,text width=2.8cm] 
    (ver) at (4.65,-1.45)
    {\textcolor{veracity}{\textbf{Veracity association}}\\
     reply states and transitions\\
     sparse LDA score};

\node[box,fill=white,text width=2.65cm] 
    (reach) at (8.5,1.45)
    {Recorded-growth forecast};

\node[box,fill=white,text width=2.65cm] 
    (evidence) at (8.5,-1.45)
    {Response explanation\\+ retrieved evidence};

\node[box,fill=softgreen,text width=2.8cm] 
    (human) at (12.5,0)
    {\textcolor{triage}{\textbf{Human reviewer}}\\
     prioritizes, checks, and decides};


\draw[arr] (prefix.east) -- (vir.west);
\draw[arr] (prefix.east) -- (ver.west);

\draw[arr] (vir.east) -- (reach.west);
\draw[arr] (ver.east) -- (evidence.west);

\draw[arr] (reach.east) -- (human.west);
\draw[arr] (evidence.east) -- (human.west);

\end{tikzpicture}

\caption{Proposed asynchronous early-triage workflow. The growth forecast is available at 30 minutes, while response context is added when 15 replies have arrived. The two predictive axes remain separate, and each contributes evidence to a human review queue.}
\label{fig:system}
\end{figure}

Our primary question is: given a propagation tree observed for 30 minutes, how much recorded growth is likely to remain? Early node count provides a strong and operationally simple baseline, but equally large snapshots can be organized differently across propagation depths and arrival times. We test whether structural and temporal entropy capture this organization and improve forecasting beyond early activity volume alone. As a complementary question, we ask whether the first 15 replies contain response patterns associated with claims later labeled true or false. These targets have different evidence requirements, so we model them separately and integrate their outputs only in the proposed human-review interface (Figure~\ref{fig:system}).

The two branches use observation boundaries suited to their tasks. A fixed 30-minute snapshot supports forecasting the recorded growth that remains after early activity has been observed. The response analysis becomes available when a thread reaches 15 chronological replies: among the 563 eligible PHEME threads, the median time to that point is 28.8 minutes (IQR, 12.9--73.4), with 52.0\% reaching it within 30 minutes and 71.6\% within one hour. The proposed architecture therefore allows a growth forecast to enter the review queue at 30 minutes and adds response context when the required reply prefix becomes available.

The paper makes three contributions:

\begin{enumerate}
    \item We evaluate whether structural and temporal entropy improve prediction of subsequent recorded FibVID propagation-tree growth beyond early node count, identify a larger improvement among high-activity trees, and quantify the improvement with paired claim-bootstrap confidence intervals.
    \item We characterize early conversational states and transitions, report event-held-out performance without explicit factual-alignment judgments as the primary response result, and retain a directly inspectable ten-term sparse linear discriminant axis for descriptive analysis.
    \item We propose a human-in-the-loop triage architecture that integrates predicted growth, response-pattern association, and retrieved evidence to prioritize claims as they enter human review queues. 
    
\end{enumerate}

\section{Related work}

\subsection{Early virality and propagation structure}

Early cascade forecasting has long used partial observations of diffusion. SEISMIC estimates infectiousness from observed reshares and forecasts final popularity \citep{zhao2015seismic}; structural virality distinguishes broadcast-like cascades from multi-generation spread \citep{goel2016structural}. Later work has shown that propagation structure and timing carry useful signals for rumor analysis \citep{zubiaga2016analysing,wei2019modeling,peng2025rumor}. Our setting differs in its observation rule and ablation target. Every modeled FibVID propagation tree is observed for 30 minutes from its root, and the central test asks whether depth-distribution and arrival-time entropy improve prediction of subsequent recorded growth beyond early activity volume.

Our ``structural entropy'' is deliberately simple: Shannon entropy over the depths occupied by nodes observed in the first 30 minutes. It is related in motivation, but not identical, to graph structural information based on coding trees \citep{li2016structural,peng2025rumor}. This distinction matters because our quantity has a direct operational interpretation---how broadly the observed activity is distributed across propagation generations---and can be computed in a streaming system without optimizing a graph partition.

\subsection{Social response as a veracity signal}

Conversation-level veracity models have used stance sequences, reply topology, and temporal dynamics \citep{kochkina2018all,wei2019modeling}. PHEME established the value of reply conversations collected around breaking-news rumors \citep{zubiaga2016analysing}. Social reactions can also differ systematically between true and false content; for example, false stories may evoke different emotional or moral responses \citep{prollochs2021emotion}. These findings motivate our use of multiple reply dimensions, but our goal is narrower than end-to-end rumor classification. We seek a compact, inspectable axis that describes how response distributions and transitions differ between labeled groups.

Information theory provides a common language for this analysis. Shannon entropy measures uncertainty in a categorical distribution \citep{shannon1948}; conditional entropy, mutual information, and pointwise mutual information describe predictability and dependence across consecutive response states. LDA then identifies a linear direction that separates two labeled distributions \citep{fisher1936}, while Cohen's $d$ expresses the separation in pooled standard-deviation units \citep{cohen1988}. The resulting score is easy to audit: every term is a named response pattern with a signed coefficient.

\subsection{Human review and evidence retrieval}

Automated fact-checking systems commonly begin by ranking check-worthy claims for limited human attention \citep{hassan2017claimbuster}. Mendes et al. evaluate a human-in-the-loop workflow for emerging COVID-19 treatment misinformation, including claim aggregation, ranking, and policy review \citep{mendes2023human}. We adopt the same operational principle: prioritization is useful even when full automation is inappropriate.

Evidence retrieval is complementary to early social-response analysis. Retrieval-augmented generation grounds downstream outputs in an external corpus \citep{lewis2020rag}; Check-COVID provides expert-annotated COVID-19 claims paired with scientific evidence \citep{wang2023checkcovid}. Our secondary retrieval study evaluates whether gold evidence is surfaced, not whether a generated verdict is correct. This boundary keeps evidence availability separate from veracity classification.

\section{Problem formulation}

For virality, let propagation tree $c$ consist of a root tweet and subsequent recorded retweet nodes ordered by timestamp, with dataset-observed terminal size $N_c$. The snapshot $P_c^{30}$ contains the nodes recorded by 30 minutes from the root and $n_c^{30}=|P_c^{30}|$. We estimate $\log(1+N_c-n_c^{30})$ using only this snapshot, so the target measures subsequent recorded propagation-tree growth.

For veracity, the fifteenth reply defines the trigger time $\tau_{15}$ and the observed prefix is
\begin{equation}
P_c = \{v_0,v_1,\ldots,v_{15}\},
\end{equation}
where $v_0$ is the root.

For the veracity task, each of the 15 replies is mapped to a collection of categorical response states. From the state sequences we calculate composition, entropy, and transition statistics. The output $D_c\in\R$ is an association score whose positive direction is aligned with PHEME's true label and whose negative direction is aligned with its false label. $D_c$ is not an evidence-grounded truth probability. In our proposed workflow, the recorded-growth forecast and $D_c$ would determine priority and explanation, while the reviewer would inspect the claim and retrieved sources before deciding what action, if any, is appropriate.

\section{Data}

\subsection{Datasets and representation}

We use separate datasets for virality, veracity, and scientific-evidence retrieval. A common relational representation retains source-specific labels, original identifiers, timing representation, text availability, and a leakage-safe \texttt{group\_id} for the propagation and reply trees.

The virality entropy ablation uses only FibVID: all 1,774 rooted propagation trees from 295 original claims \citep{triage_fibvid_2021}. These trees represent tweet and retweet propagation, not reply conversations; node count includes the root tweet and recorded retweet nodes. Native timestamps and recorded propagation depth support a 30-minute observation window without a minimum early node-count requirement. Trees from one claim share a split group, preventing claim fragments from crossing the fitting and evaluation partitions. PHEME is used only for veracity and is not pooled into the virality experiment.

The veracity analysis uses PHEME rumors with a binary true/false label and at least 15 raw chronological replies. The final sample contains 563 threads from six events, comprising 414 true and 149 false claims. Event counts are shown in Appendix~\ref{app:data}. Grouping by event creates a demanding evaluation: the model must transfer to a breaking-news context that was absent from training. Within this eligible cohort, the 15th reply arrives after a median of 28.8 minutes (IQR, 12.9--73.4); 52.0\% of threads reach that point within 30 minutes and 71.6\% within one hour.

For scientific evidence retrieval, we use the official Check-COVID test split: 229 evaluated claims, a corpus of 347 documents, and 3,459 candidate sentences. Gold evidence identifiers are used only after retrieval for evaluation.

\begin{table}[t]
\centering
\caption{Datasets used in the reported experiments. ``Eligible'' indicates that all task-specific observation and label requirements were met.}
\label{tab:data}
\small
\begin{tabularx}{\textwidth}{@{}l l r Y Y@{}}
\toprule
Branch & Source & $n$ & Unit & Evaluation role \\
\midrule
Virality & FibVID & 1,774 & rooted propagation trees from 295 claims & 1,422 training; 352 test trees, split by claim \\
Veracity & PHEME & 563 & binary-labeled threads with $\geq 15$ raw replies & event-held-out screen; descriptive sparse LDA \\
Retrieval & Check-COVID & 229 & official test claims & gold-evidence retrieval evaluation \\
\bottomrule
\end{tabularx}
\end{table}

\subsection{Reply-state annotation}

Each PHEME reply was labeled with Qwen2.5-7B-Instruct in 4-bit inference on an NVIDIA A10G \citep{triage_qwen25_2024}. The full annotation prompt is reproduced in Appendix~\ref{app:qwen-prompt}. The structured annotation schema contains nine dimensions:

\begin{enumerate}
    \item Stance toward the root claim
    \item Stance toward the parent reply
    \item Communicative act
    \item Factual alignment
    \item Primary emotion
    \item Emotion valence
    \item Emotion intensity
    \item Emotion target
    \item Affective claim
\end{enumerate}

The run produced 8,445 expected reply positions across 563 threads. Of these, 8,397 (99.43\%) passed parsing and schema validation, while 48 (0.57\%) remained invalid and were retained as missing at their original reply positions. No later replies were used to replace invalid annotations. The model-generated labels are measurement inputs, not human-standard annotations, acknowledged in Section~\ref{sec:limitations}.

\section{Methods}
\subsection{Virality features}

The reported virality ablation uses FibVID's 1,774 rooted propagation trees from 295 original claims. We observe each tree for 30 minutes from its root, retaining root-only trees and trees with no subsequent recorded growth. Let $t_i$ be node $i$'s elapsed arrival time in seconds and define the snapshot
\begin{equation}
P_c^{30}=\{v_i:0\leq t_i\leq 1800\},\qquad n_c^{30}=|P_c^{30}|.
\end{equation}
The observed count $n_c^{30}$ includes the tree root. If $N_c$ is the terminal number of nodes recorded in the dataset, the prediction target is
\begin{equation}
G_c=N_c-n_c^{30},\qquad Y_c=\log(1+G_c).
\label{eq:futuregrowth}
\end{equation}
Here $\log$ denotes the natural logarithm. The outcome excludes nodes already observed at the cutoff. All predictors use only the 30-minute snapshot; terminal size is used only to construct the outcome. This virality observation rule is separate from the first-15-reply boundary used for PHEME veracity analysis.

\begin{figure}[H]
    \centering

\resizebox{0.55\linewidth}{!}{%
\begin{tikzpicture}[
    >=Latex,
    baseNode/.style={
        circle,
        draw=black,
        line width=1pt,
        minimum size=10mm,
        inner sep=0pt
    },
    edge/.style={
        ->,
        line width=1pt
    },
    bucketline/.style={
        dashed,
        gray!70,
        line width=0.8pt
    },
    depth0/.style={baseNode, fill=depth0fill, text=white},
    depth1/.style={baseNode, fill=depth1fill, text=white},
    depth2/.style={baseNode, fill=depth2fill},
    depth3/.style={baseNode, fill=depth3fill}
]

\def\xzero{0}
\def\xone{3}
\def\xtwo{6}
\def\xthree{9}
\def\xfour{12}

\def\ytop{5.7}
\def\ybottom{-0.5}

\node[font=\bfseries\Large] at (6,7.1) {Time Buckets};

\node[font=\large] at (1.5,6.45) {$t=0\,\mathrm{min}$};

\draw[
    decorate,
    decoration={brace,amplitude=5pt}
]
(\xzero,5.95) -- (\xone,5.95);

\node[font=\large] at (6,6.45) {10--20 min};

\draw[
    decorate,
    decoration={brace,amplitude=5pt}
]
(\xone,5.95) -- (\xthree,5.95);

\node[font=\large] at (10.5,6.45) {20--30 min};

\draw[
    decorate,
    decoration={brace,amplitude=5pt}
]
(\xthree,5.95) -- (\xfour,5.95);

\foreach \x in {\xzero,\xone,\xtwo,\xthree,\xfour} {
    \draw[bucketline] (\x,\ytop) -- (\x,\ybottom);
}

\node[depth0, font=\bfseries] (S) at (1.35,3.0) {S};

\node[depth1] (A) at (4.5,4.85) {};
\node[depth1] (B) at (4.5,3.0) {};
\node[depth1] (C) at (4.5,1.25) {};

\node[depth2] (D) at (7.5,4.85) {};
\node[depth2] (E) at (7.5,3.0) {};

\node[depth3] (F) at (10.5,3.0) {};

\draw[edge] (S) -- (A);
\draw[edge] (S) -- (B);
\draw[edge] (S) -- (C);

\draw[edge] (A) -- (D);

\draw[edge] (B) -- (E);
\draw[edge] (E) -- (F);

\node[font=\large] at (1.35,-1.35) {$d=0$};
\node[font=\large] at (4.5,-1.35) {$d=1$};
\node[font=\large] at (7.5,-1.35) {$d=2$};
\node[font=\large] at (10.5,-1.35) {$d=3$};

\node[font=\bfseries\large] at (6,-2.1) {Depth ($d$)};

\end{tikzpicture}%
}

\caption{Illustration of structural depth and temporal bucketing in an early propagation tree.}
\label{fig:time-buckets}

\end{figure}
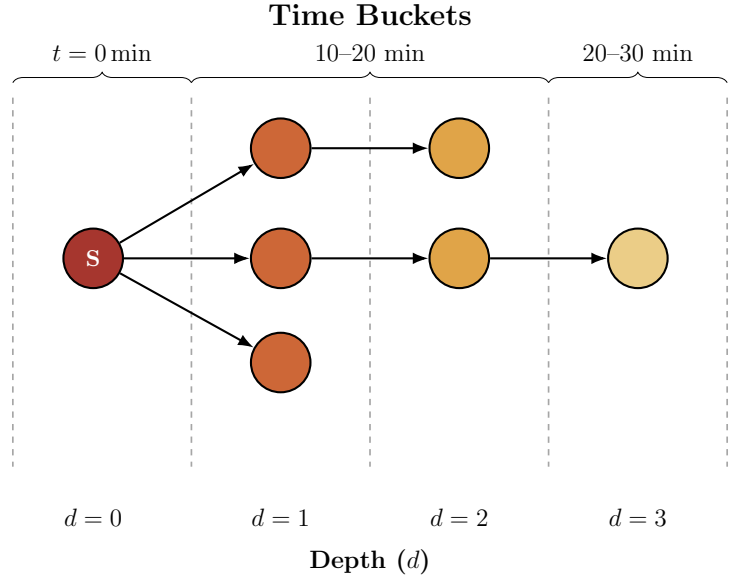

\paragraph{Structural entropy.}
Let $d_i$ be the propagation-tree depth of node $v_i$ in $P_c^{30}$, and let $p_k$ be the fraction of observed nodes at depth $k$, including the root. We define
\begin{equation}
H_S(P_c^{30})=-\sum_k p_k\log_2 p_k.
\label{eq:structentropy}
\end{equation}
A star-shaped propagation tree concentrates nearly all non-root nodes at depth one and has low depth diversity. A tree whose early activity occupies several generations has higher $H_S$. We use the unnormalized entropy in bits and adopt $0\log_2 0=0$; a root-only snapshot has $H_S=0$.

\paragraph{Temporal entropy.}
For observed nodes with $0<t_i\leq1800$, let $b_i=\min\{5,\lfloor t_i/300\rfloor\}$ identify one of six fixed five-minute arrival bins. The last bin includes the 30-minute endpoint. Let $q_b$ be the fraction of these positive-time arrivals in bin $b$. We define
\begin{equation}
H_T(P_c^{30})=-\sum_{b=0}^{5} q_b\log_2 q_b.
\label{eq:tempentropy}
\end{equation}
Low temporal entropy indicates that activity is concentrated in a small number of time bins; high entropy indicates a more even distribution across the window. The root and any arrivals recorded at zero elapsed time are excluded from this distribution. We use unnormalized entropy in bits and set $H_T=0$ when no positive-time arrival is observed.

\paragraph{Baseline and ablation.}
The baseline uses only $n_c^{30}$, a direct measure of early activity volume. Three ablations add $H_S$, $H_T$, or both. Each configuration uses XGBoost squared-error regression with 200 boosting rounds, maximum depth three, learning rate 0.04, minimum child weight 10, $L_2$ regularization 2, row and feature subsampling fractions of 0.8, and random seed 20260917 \citep{chen2016xgboost}. These common settings isolate the added predictive value of the entropy features.

We partition by original claim using GroupShuffleSplit with a 20\% test-group fraction and seed 20260915. All rooted propagation trees of one claim remain in the same partition: 1,422 trees from 236 claims are used for fitting and 352 trees from 59 claims for scoring. We report MAE and $R^2$ on $Y_c$ and Spearman rank correlation between observed and predicted $Y_c$. We also report future-node MAE after converting predictions to $\widehat G_c=\exp(\max\{0,\widehat Y_c\})-1$. An eventual recorded-size forecast can be reconstructed as $n_c^{30}+\widehat G_c$.

To examine performance among already-active propagation trees, we additionally score the same fitted models on test trees with $n_c^{30}\geq15$. This cutoff is approximately the upper quartile of 30-minute activity in FibVID and includes 97 test trees from 35 claims. The subgroup is defined entirely by information available at the observation cutoff; models are fitted on all training trees and are not refitted for the subgroup. The four model configurations, fitting settings, and subgroup rule are fixed for the reported evaluation. We treat the ablation as an internal exploratory analysis with separate claim groups at fitting and scoring.

We quantify uncertainty in the improvement from count alone to count plus both entropies using a paired claim-cluster bootstrap with 10,000 iterations and seed 20261002. Each iteration samples original test-set claim groups with replacement and retains every propagation tree belonging to each sampled claim, preserving the paired predictions and within-claim dependence. We report percentile 95\% confidence intervals for the change in $R^2$ and the relative reduction in log-MAE. The intervals quantify variation across test claims conditional on the fitted models.

\subsection{Veracity feature construction}

For each response-label dimension in the annotation schema (Appendix~\ref{app:qwen-prompt}), the first 15 raw chronological replies define fixed positions
$X_1,\ldots,X_{15}$. If the annotation at position $i$ is invalid, $X_i$ is treated as missing and its original reply position is preserved; later replies are never used as replacements. Let
\[
\mathcal{V}=\{i\in\{1,\ldots,15\}:X_i\text{ is valid}\}
\]
denote the set of validly annotated positions.

For state $a$, we compute its ordinary share
\[
\hat p(a)
=
\frac{\sum_{i\in\mathcal{V}}\mathbf{1}(X_i=a)}
{|\mathcal{V}|},
\]
and its depth-weighted share
\[
\hat p_{\mathrm{depth}}(a)
=
\frac{\sum_{i\in\mathcal{V}} d_i\,\mathbf{1}(X_i=a)}
{\sum_{i\in\mathcal{V}} d_i},
\]
where $d_i$ is the reply-tree depth of reply $i$. The ordinary share measures the prevalence of a response state within the prefix, whereas the depth-weighted share gives greater influence to states occurring in deeper, more nested parts of the reply tree.

We additionally compute (an empirical) Shannon and normalized entropy, dominant-state share, number of observed states, entropy over the early and late portions of the prefix, entropy change, and the rate at which adjacent states switch. Early and late quantities are defined by the original reply ranks 1--7 and 8--15, respectively, so missing annotations do not shift replies between portions of the prefix.

For transition-based features, define the eligible adjacency set
\[
\mathcal{E}
=
\left\{
i\in\{1,\ldots,14\}:
X_i \text{ and } X_{i+1} \text{ are both valid}
\right\},
\]
and let $T=|\mathcal{E}|.
$
For states $(a,b)$, let
\[
n(a,b)
=
\sum_{i\in\mathcal{E}}
\mathbf{1}\!\left(X_i=a,\;X_{i+1}=b\right)
\]
denote the number of observed $a\!\to\!b$ transitions between consecutive original reply ranks. Three feature types require careful distinction:
\begin{align}
\text{motif}(a\!\to\!b)
&=
\frac{n(a,b)}{T},
\\
\widehat P(b\mid a)
&=
\frac{n(a,b)}
{\sum_{b'}n(a,b')},
\\
\widehat P(a,b)
&=
\frac{n(a,b)}{T},
\\
\operatorname{PMI}(a,b)
&=
\log_2
\frac{\widehat P(a,b)}
{\widehat P_{\mathrm{src}}(a)\widehat P_{\mathrm{dst}}(b)},
\end{align}
where
\[
\widehat P_{\mathrm{src}}(a)
=
\frac{\sum_{b'}n(a,b')}{T},
\text{ }
\widehat P_{\mathrm{dst}}(b)
=
\frac{\sum_{a'}n(a',b)}{T}.
\]

The motif share measures how much of the eligible transition stream is occupied by a particular pair; the conditional transition probability measures what follows state $a$ when $a$ appears as the source of an eligible transition; and PMI measures whether the pair occurs more often than expected from its source- and destination-state marginal frequencies. If an annotation at position $i$ is missing, no transition is formed across that position; in particular, $X_{i-1}\!\to\!X_{i+1}$ is never counted as an observed transition.

We additionally calculate transition entropy, conditional entropy, and mutual information over the eligible transition set. Prefix-only reply-tree timing and structure and basic lexical entropy are available for the broader feature-family comparison. Completed-thread quantities are never imported into the prefix features.

\subsection{Event-held-out feature-family screen}

We compare five prespecified feature families using five-fold
StratifiedGroupKFold, with complete events held out between training
and evaluation. The primary screen excludes every feature derived from the
explicit factual-alignment judgment, while retaining stance toward the root,
stance toward the parent, communicative act, and the five affective dimensions.
This removes the direct LLM-generated accuracy judgment without removing the
separate textual indicators of endorsement or contradiction. \emph{Prefix structure and timing} contains eight candidate
variables describing prefix size, reply-tree depth and branching, degree
entropy, average response time, and response rate. \emph{State composition}
contains 94 ordinary and depth-weighted state-prevalence variables.
\emph{Entropy and transition dynamics} contains 1,113 variables describing state uncertainty, dominance, diversity, early--late entropy change, switching, transition entropy, conditional entropy, mutual information, pairwise motif shares, conditional transition probabilities, and PMI. \emph{All response and lexical features} contains these response-derived variables together with the 94 state-composition variables and two basic lexical-diversity measures, for 1,209 candidate variables. \emph{Combined response, lexical, and structural} contains all 1,217 candidate variables. Counts are reported before within-fold variance filtering and sparse regularization. Within each training fold, missing values are
median-imputed, constant features are removed, remaining features are
standardized, and an L1-regularized, class-weighted logistic-regression model
is fitted using a prespecified regularization value of $C=0.1$. We pool
out-of-fold probabilities and report ROC--AUC, area under the
precision--recall curve (PR--AUC), balanced accuracy, and macro-F1 at a
probability threshold of 0.5.

As a matched sensitivity comparison, we rerun the same folds, preprocessing,
model specification, and threshold with the factual-alignment-derived features
included. The complete fixed-prefix matrix contains 97 such columns; after
the already-prespecified exclusion of raw annotation-count fields, they add
10 state-composition candidates, 86 entropy-and-dynamics candidates, and 96
candidates to each broad response family. Prefix structure and timing contain
no factual-alignment feature and therefore provide an unchanged pipeline control.

\subsection{Sparse LDA axis}

As a secondary descriptive analysis, the sparse axis retains the original
nine-dimension annotation representation, including factual alignment. To
obtain a compact descriptive axis, we begin with 1,094 nonconstant
label-derived candidates from the fixed-prefix feature matrix.
Annotation-coverage fields and raw annotation-count variables are excluded.
We median-impute and standardize the remaining features, use recursive feature
elimination with an $\ell_1$-regularized, class-weighted logistic-regression
estimator to retain ten variables, and fit one-dimensional LDA to the selected
matrix. If $z_{cj}$ denotes standardized feature $j$ for thread $c$, the
resulting axis is

\[
D_c=\sum_{j=1}^{10}\beta_j z_{cj}.
\]

The orientation is chosen so that larger scores correspond to the
true-labeled side of the fitted axis. We quantify separation between the two
projected distributions using Cohen's $d$ and the two-sample KS statistic.

This sparse analysis is explicitly descriptive: feature selection and LDA
fitting use the same 563-thread cohort on which separation is measured.
Accordingly, its effect size and coefficients characterize the observed
sample rather than transfer to unseen events. The event-held-out
feature-family experiment provides the paper's evidence about classification
performance.

\subsection{Secondary scientific-evidence retrieval}

The retrieval pipeline takes claim text as its only query input. BM25 and all-MiniLM-L6-v2 dense retrieval independently rank documents; reciprocal rank fusion (RRF, $k=60$) combines their rankings \citep{cormack2009rrf,reimers2019sbert}. From the top 20 fused documents, a cross-encoder (ms-marco-MiniLM-L-6-v2) reranks sentences. We report recall at $k$ and mean reciprocal rank (MRR) against Check-COVID's gold evidence. No retriever or reranker is fine-tuned on the test split.

\section{Results}

\subsection{Entropy improves continuation prediction beyond early count}

Table~\ref{tab:viralityall} reports the four-model ablation on 352 FibVID test propagation trees from 59 claim groups separate from training. The count-only baseline reaches $R^2=0.3067$ and Spearman's $\rho=0.5401$ for post-30-minute recorded growth. Adding structural entropy raises these values to 0.3196 and 0.5514. Adding temporal entropy alone yields $R^2=0.3039$ and $\rho=0.5401$, with no improvement in explained variation. The three-feature model containing count and both entropies performs best among the four configurations and improves every reported metric: $R^2=0.3227$, $\rho=0.5535$, and log-MAE 1.2029. Relative to count alone, it gains 0.0161 in $R^2$ (95\% claim-bootstrap CI, $-0.0064$ to 0.0440) and reduces log-MAE by 2.4\% (95\% CI, $-0.4$\% to 5.2\%). Future-node MAE falls from 75.60 to 72.21.

\begin{table}[ht]
\centering
\caption{FibVID propagation-tree continuation ablation on all 352 test trees from 59 claim groups. Log-MAE and $R^2$ use $\log(1+G_c)$; node MAE uses the reconstructed number of subsequent recorded nodes. Errors are lower-is-better. Bold denotes the best value among the four configurations.}
\label{tab:viralityall}
\small
\begin{tabularx}{\textwidth}{@{}Yrrrr@{}}
\toprule
Model & Log-MAE & $R^2$ & Spearman $\rho$ & Node MAE \\
\midrule
Count only & 1.2323 & 0.3067 & 0.5401 & 75.60 \\
Count + structural entropy & 1.2139 & 0.3196 & 0.5514 & 74.10 \\
Count + temporal entropy & 1.2334 & 0.3039 & 0.5401 & 75.61 \\
Count + both entropies & \textbf{1.2029} & \textbf{0.3227} & \textbf{0.5535} & \textbf{72.21} \\
\bottomrule
\end{tabularx}
\end{table}

\subsection{The gain is larger among high-activity propagation trees}

The high-activity subgroup contains 97 test propagation trees from 35 claims with at least 15 nodes observed by 30 minutes. Table~\ref{tab:viralityhigh} scores the same fitted models on this subgroup. Count alone reaches $R^2=0.2482$ and $\rho=0.3941$. Adding structural entropy raises $R^2$ to 0.3738 and $\rho$ to 0.5239. Adding temporal entropy without structural entropy lowers log-MAE from 0.6400 to 0.6253 and raises $\rho$ to 0.4327, but lowers $R^2$ to 0.2396.

Adding both entropies raises $R^2$ to 0.3954, a gain of 0.1472 over count alone (95\% claim-bootstrap CI, $-0.0318$ to 0.3076), and raises $\rho$ to 0.5288, a gain of 0.1347. Log-MAE falls from 0.6400 to 0.5651, an 11.7\% reduction (95\% CI, $-1.9$\% to 23.9\%), while future-node MAE falls from 76.17 to 67.22. Relative to count plus structural entropy, adding temporal entropy raises $R^2$ by a further 0.0216 and reduces log-MAE from 0.5882 to 0.5651. The point estimates favor the combined model across every reported metric. Appendix~\ref{app:virality} shows the high-activity predictions against observed subsequent growth.

\begin{table}[ht]
\centering
\caption{FibVID propagation-tree continuation ablation on 97 high-activity test trees from 35 claims, defined by $n_c^{30}\geq15$. Each model is fitted on the same complete training partition used in Table~\ref{tab:viralityall}. The subgroup is selected using only 30-minute activity.}
\label{tab:viralityhigh}
\small
\begin{tabularx}{\textwidth}{@{}Yrrrr@{}}
\toprule
Model & Log-MAE & $R^2$ & Spearman $\rho$ & Node MAE \\
\midrule
Count only & 0.6400 & 0.2482 & 0.3941 & 76.17 \\
Count + structural entropy & 0.5882 & 0.3738 & 0.5239 & 72.94 \\
Count + temporal entropy & 0.6253 & 0.2396 & 0.4327 & 74.31 \\
Count + both entropies & \textbf{0.5651} & \textbf{0.3954} & \textbf{0.5288} & \textbf{67.22} \\
\bottomrule
\end{tabularx}
\end{table}

\FloatBarrier

\subsection{Early response-state composition transfers across held-out events}

After excluding every explicit factual-alignment-derived feature, state
composition provides the strongest ranking performance among the
response-derived families, with event-held-out ROC--AUC 0.538 and PR--AUC
0.764 (Table~\ref{tab:veracityscreen}). Its balanced accuracy is 0.522 and its
macro-F1 is 0.484. The combined response, lexical, and structural family
achieves ROC--AUC 0.492 and PR--AUC 0.745. All response and lexical features
yield ROC--AUC 0.488 and PR--AUC 0.742. Prefix structure and timing alone,
which are unchanged by the exclusion, yield ROC--AUC 0.539 and macro-F1
0.506, whereas entropy and transition dynamics yield ROC--AUC 0.460 and
macro-F1 0.466.

Among the response-derived families, the strongest cross-event signal comes
from the ordinary and depth-weighted prevalence of stance, communicative, and
affective states. The substantially larger entropy and transition-dynamics
family does not improve ranking performance beyond state composition alone.
PR--AUC should also be interpreted relative to the 73.5\% prevalence of
true-labeled threads. Balanced accuracy and macro-F1 remain close to 0.5 at
the prespecified 0.5 probability threshold.

Performance also varies across held-out events. For the combined model
without factual alignment, ROC--AUC is 0.568 for Sydney Siege, 0.474 for
Ottawa Shooting, 0.426 for Germanwings, 0.459 for Charlie Hebdo, and 0.857 for the combined
Ferguson/Gurlitt fold. The Ferguson/Gurlitt fold contains only ten threads,
including three false and seven true threads, so its high AUC is highly
unstable and should not be interpreted independently. The aggregate
ROC--AUC of 0.492 is calculated from the pooled out-of-fold probabilities
rather than by averaging the five fold-specific AUCs.

\begin{table}[t]
\centering
\caption{Primary PHEME event-grouped out-of-fold evaluation after excluding explicit factual-alignment-derived features. Each feature-family model is trained without the held-out events. Candidate counts are reported before within-fold variance filtering and sparse regularization. PR--AUC should be interpreted relative to the 73.5\% true-class prevalence.}
\label{tab:veracityscreen}
\small
\setlength{\tabcolsep}{4pt}
\begin{tabularx}{\textwidth}{@{}Yrrrrr@{}}
\toprule
Feature family & Candidates & ROC--AUC & PR--AUC & Bal.\ acc. & Macro-F1 \\
\midrule
Combined response, lexical, and structural
    & 1,217 & 0.492 & 0.745 & 0.471 & 0.465 \\
State composition
    & 94 & 0.538 & 0.764 & 0.522 & 0.484 \\
All response and lexical features
    & 1,209 & 0.488 & 0.742 & 0.467 & 0.463 \\
Prefix structure and timing
    & 8 & \textbf{0.539} & \textbf{0.765} & \textbf{0.522} & \textbf{0.506} \\
Entropy and transition dynamics
    & 1,113 & 0.460 & 0.720 & 0.473 & 0.466 \\
\bottomrule
\end{tabularx}
\end{table}

A secondary comparison that includes the explicit factual-alignment features
quantifies how much this LLM-generated accuracy judgment contributes. State-composition
ROC--AUC rises from 0.538 to 0.562 and PR--AUC from 0.764 to 0.774. The broader
response and lexical family rises from 0.488 to 0.543, and the combined family
from 0.492 to 0.546. The unchanged structural family reproduces ROC--AUC 0.539
in both runs. This comparison shows that factual alignment contributes to the
full representation while preserving a separately reported response-state
result that does not use the explicit accuracy judgment.

\subsection{The sparse LDA axis yields an interpretable separation}

\begin{figure}[t]
\centering
\includegraphics[width=0.82\textwidth]{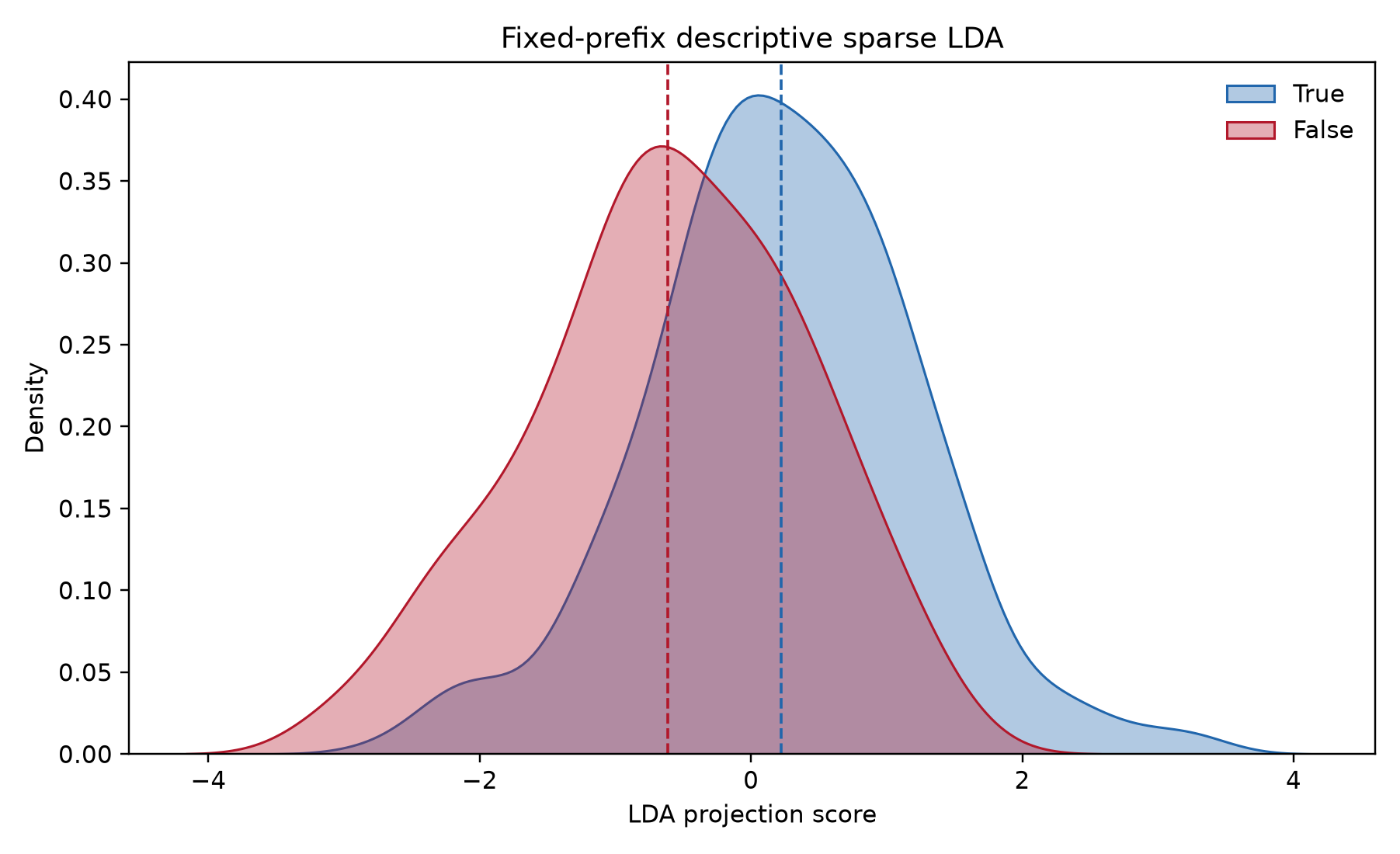}
\caption{Density of the 563 eligible PHEME threads projected onto the descriptive ten-feature sparse LDA axis. Positive values are oriented toward the true-associated distribution. Feature selection and fitting use this same complete sample, so the figure illustrates descriptive separation rather than held-out classification.}
\label{fig:lda}
\end{figure}

Retaining the complete nine-dimension representation for this secondary
descriptive fit, RFE selected ten variables from 1,094 nonconstant label-derived candidates,
after which LDA constructed a one-dimensional axis using the same complete
563-thread cohort. The resulting projection separates the true- and
false-labeled score distributions with Cohen's $d=0.8321$ and KS statistic
$D_{\mathrm{KS}}=0.3523$ (Figure~\ref{fig:lda}). These are descriptive
full-sample statistics calculated after feature selection and fitting; they
are not estimates of performance on unseen events.

Table~\ref{tab:ldacoef} gives the complete descriptive axis. The variables are
standardized before multiplication, and each coefficient represents a
conditional contribution given the other nine selected variables. Positive
coefficients move the fitted score toward the true-labeled side, while
negative coefficients move it toward the false-labeled side. The largest
positive coefficient is the PMI of the
blame/accusation$\to$threat/harm transition ($+0.5885$). The largest negative
coefficients are the motif share of inaccurate$\to$no factual claim
($-0.5586$) and
$P(\text{sadness}\mid\text{fear/anxiety})$ ($-0.5437$).

\begin{center}
\centering
\begin{minipage}{\textwidth}
\captionsetup{hypcap=false}
\captionof{table}{Descriptive sparse-LDA axis fitted to all 563 eligible PHEME threads. RFE selected ten variables from 1,094 nonconstant label-derived candidates. Each variable was standardized before multiplication. Positive coefficients move the score toward the true-labeled side of the fitted axis; negative coefficients move it toward the false-labeled side. These coefficients explain the observed full-sample separation of Cohen's $d=0.832$ rather than held-out classification performance.}
\label{tab:ldacoef}
\small
\begin{tabularx}{\textwidth}{@{}Y Y r@{}}
\toprule
Response dimension & Standardized feature & $\beta$ \\
\midrule
Affective claim
& PMI: blame/accusation $\to$ threat/harm
& $+0.5885$ \\

Factual alignment
& motif share: inaccurate $\to$ no factual claim
& $-0.5586$ \\

Primary emotion
& $P(\text{sadness}\mid\text{fear/anxiety})$
& $-0.5437$ \\

Emotion valence
& late-prefix entropy
& $+0.5225$ \\

Affective claim
& motif share: blame/accusation $\to$ threat/harm
& $-0.4858$ \\

Primary emotion
& PMI: fear/anxiety $\to$ sadness
& $+0.4486$ \\

Emotion valence
& conditional transition entropy
& $-0.4200$ \\

Factual alignment
& motif share: inaccurate $\to$ accurate
& $-0.4020$ \\

Emotion valence
& depth-weighted share: negative
& $+0.3877$ \\

Affective claim
& PMI: feeling expression $\to$ hope/reassurance
& $+0.1964$ \\
\bottomrule
\end{tabularx}
\end{minipage}
\end{center}

\subsection{Scientific retrieval can support the reviewer}

RRF retrieves a gold evidence document for 74.2\% of Check-COVID test claims within the top five documents and 94.3\% within the top 20, with MRR 0.563. Dense retrieval and BM25 individually reach Recall@20 of 89.1\% and 83.0\%, respectively. After candidate generation, the cross-encoder retrieves a gold sentence within the top five for 38.4\% of claims and within the top 20 for 58.1\%, with MRR 0.295. The document-stage result provides a useful evidence pool; sentence selection remains the main retrieval bottleneck. These figures measure retrieval only and should not be reported as fact-checking accuracy.

\section{Discussion}

\subsection{When organization adds information beyond volume}

Early node count is a defensible forecasting baseline: it measures how much activity a propagation tree has accumulated in a fixed window. The continuation target then asks how much additional recorded activity remains, excluding the nodes already observed. In this setting, the entropy ablation tests whether the organization of early activity contributes beyond its volume. Structural entropy summarizes the distribution of observed activity across propagation depths, while temporal entropy summarizes its concentration across fixed arrival-time bins. Propagation trees with the same early count can have different values of both quantities.

The combined entropy model improves every reported metric across all test propagation trees, with especially pronounced gains among trees already showing substantial activity. In the high-activity subgroup, adding the entropy pair increases $R^2$ by 0.1472 and reduces log-MAE by 11.7\% relative to count alone. This pattern is consistent with organization becoming more informative once enough nodes are present to distinguish depth diversity and temporal concentration. The subgroup analysis scores the same models rather than fitting a specialized high-activity model, so it describes where the added features improve the count-based forecast.

Structural entropy accounts for most of the predictive improvement. Adding temporal entropy to count without structural entropy does not improve $R^2$ in either test population, although it improves high-activity rank correlation and absolute error. Its contribution is clearest when paired with structural entropy: the combined model improves on count plus structural entropy in both populations. These results support a compact characterization of where and when activity occurs, without establishing that either entropy causes subsequent growth or that the three-feature model outperforms every possible forecasting feature set.

The ablation assesses propagation trees from separate original claims within one corpus. Replication on independent claims, later periods, and other platforms is needed to establish transfer. Recorded propagation-tree growth also differs from audience exposure: the outcome counts nodes preserved in FibVID rather than everyone who encountered a claim.

Future work should extend depth and arrival entropy with information-theoretic metrics of branching, cross-depth dependence, and time-conditioned structure. Fixed ablations on independent claims and later periods should test whether these metrics contribute complementary, approximately orthogonal information beyond early count and the existing entropies.

\subsection{Interpreting the veracity axis and its limits}

The sparse axis provides an inspectable summary of how early-response patterns
differ within the fitted PHEME cohort. Its selected terms span
factual-alignment transitions, affective transitions, and the distribution
and sequential organization of emotional valence. A reviewer can therefore
inspect which response patterns move a thread toward either side of the
fitted axis. Because the coefficients are jointly estimated, however, their
directions describe conditional associations rather than independent or
causal effects.

Two transitions enter the fitted axis through more than one statistic. For
the transition from blame/accusation to threat/harm, the motif coefficient is
negative while the PMI coefficient is positive. For the transition from
fear/anxiety to sadness, the conditional
transition-probability coefficient is negative while the PMI coefficient is
positive. These signs are not contradictory. Motif share measures the
absolute prevalence of a pair within the eligible transition stream;
conditional probability measures what follows a given source state; and PMI
measures overrepresentation relative to the source- and destination-state
marginals. A pair can occupy a small share of the complete transition stream
while still occurring more often than expected from two individually
infrequent states.

The descriptive sparse axis includes two transitions from the factual-alignment
dimension: inaccurate to no factual claim and inaccurate to accurate. This
dimension is an LLM-generated accuracy judgment and is distinct from the root-
and parent-stance dimensions that encode endorsement or contradiction. Larger
motif shares for both transitions move the fitted score toward the false-labeled
side after the remaining selected variables are held fixed. Because the primary
event-held-out result excludes factual alignment, the cross-event response claim
does not depend on these explicit accuracy judgments. The emotion-valence terms
distinguish late-prefix diversity, conditional transition uncertainty, and the
concentration of negative valence at greater reply depth. Their differing
coefficients indicate that the fitted axis uses both where emotional responses
occur and how their states change across adjacent reply positions.

The gap between descriptive separation and held-out classification is central
to the interpretation. Full-sample feature selection and LDA fitting produce
Cohen's $d=0.832$, whereas factual-alignment-free state composition reaches
event-held-out ROC--AUC 0.538. Including the explicit factual-alignment channel
in the secondary comparison raises ROC--AUC to 0.562. State composition ranks more
effectively across the evaluated held-out events than the larger transition
representation, and performance varies by event.

The sparse axis should therefore be understood as an exploratory description
of veracity-associated social responses rather than as a stand-alone truth
classifier. Its role is to expose the early-response patterns associated with
the true- and false-labeled sides of the fitted axis, providing an
interpretable signal for human review. Prospective evaluation on newly
collected events is necessary before the fixed ten-term formula can be treated
as a deployable classifier.

The observed full-sample separation motivates exploring additional information-theoretic response features, richer categories within each labeling dimension, alternative annotation prompts and models, and different classification algorithms. Human annotation audits and event- or time-held-out evaluation should test whether these extensions improve classification accuracy, rather than descriptive separation alone.

\subsection{Triage policy and interface}

Our proposed queue preserves the two axes rather than collapsing them prematurely. High predicted continuation and a strong false-associated response score are natural candidates for rapid review. A large recorded-growth forecast with an ambiguous response score can still warrant attention because discussion may expand before a claim is resolved. A low growth forecast should reduce priority only relative to competing items; it should not certify safety. Each queue item should expose the observation time, early node count, structural and temporal entropy, predicted subsequent recorded growth, signed LDA contributions, annotation coverage, and retrieved scientific passages.

Human review remains responsible for checking sources, resolving context, and applying policy. This division of labor matches the strengths of the models: rapid ranking and structured summaries at scale, followed by judgment where errors have consequences. The system can also abstain when the reply-label parse rate is poor, the thread is unlike the training events, or retrieval fails to surface relevant evidence.

\section{Limitations and responsible use}
\label{sec:limitations}

\paragraph{Conditional observation population.}
The virality analysis includes all recorded FibVID rooted propagation trees at a
30-minute observation boundary, including root-only trees and those with
no subsequent recorded growth. Its high-activity analysis additionally
conditions on at least 15 nodes observed by that cutoff. The larger subgroup
gain should therefore not be generalized to newly posted or low-activity
trees. Terminal size is the dataset-recorded endpoint, with no common
fixed future observation horizon. The veracity
results apply only to PHEME threads that contain at least 15 raw chronological
replies and have a binary true/false label.

\paragraph{Dataset and platform scope.}
The virality experiment uses only FibVID propagation trees, while the veracity
experiment covers six PHEME events. Platform behavior, moderation practices,
and retweet or reply conventions can vary across platforms and over time. Event-held-out
validation tests transfer across PHEME events, but it does not establish
prospective transfer to a future period or a different platform.
The virality claim-group split evaluates propagation trees from separate original
claims within FibVID; it does not test cross-platform transfer. This source-specific design narrows the
external scope of the results.

\paragraph{Generated reply labels.}
The nine response dimensions are generated by a 7B instruction-tuned model
rather than established through independent human annotation. Of the 8,445
fixed-prefix reply positions, 8,397 (99.43\%) pass parsing and schema
validation. Syntactic validity does not establish semantic accuracy or equal
performance across topics, dialects, and communities. Invalid annotations are
retained as missing at their original reply positions and do not create
transitions between nonadjacent replies, but validly parsed annotations may
still contain substantive labeling errors. The primary event-held-out screen
therefore excludes the explicit factual-alignment dimension, while the matched
secondary comparison quantifies its contribution. Human audit, inter-annotator
comparison, and robustness testing with alternative labelers are required
before deployment.

\paragraph{Sparse-LDA selection and generalization.}
The reported Cohen's $d=0.832$ is calculated after feature selection and LDA
fitting on the same 563-thread cohort. It is therefore a descriptive
full-sample effect size rather than an unbiased estimate of predictive
performance. The event-held-out feature-family screen evaluates
classification transfer separately and reaches ROC--AUC 0.538 without
explicit factual-alignment features and 0.562 in the secondary comparison that includes them; neither
result validates the final ten-term full-sample axis. A definitive
prospective test requires freezing the annotation schema, missing-value
handling, standardization parameters, selected variables, LDA coefficients,
and any decision rule before evaluating the system on temporally later,
independent events.

\paragraph{Association, evidence, and truth.}
Reply patterns can reflect audience composition, coordination, topic, news
context, or platform behavior. They do not establish that a claim is true or
false. Likewise, retrieval recall shows whether annotated evidence is surfaced
for a reviewer; it does not show that a downstream system interprets that
evidence correctly. The interface should distinguish these outputs explicitly
as recorded-growth forecasts, social-response associations, and retrieved evidence.

\paragraph{Potential misuse.}
Ranking can concentrate scrutiny on particular communities or allow
coordinated participants to manipulate the observed response patterns. Review
policies should audit group-level errors, record reviewer overrides, limit
automated enforcement, and make clear that absence from the review queue is
not evidence of truth, safety, or low harm.

\section{Conclusion}

Forecasting subsequent recorded propagation growth is the central contribution
of this study. In FibVID propagation trees, structural and temporal entropy
improve every reported metric after a 30-minute snapshot beyond observed node
count. The gain is especially pronounced among high-activity trees: the entropy
pair raises $R^2$ from 0.248 to 0.395 and reduces log-MAE by 11.7\% (95\%
claim-bootstrap CI, $-1.9$\% to 23.9\%). The first 15 replies provide a
complementary response-pattern signal. After explicit factual-alignment
judgments are excluded, state composition reaches event-held-out ROC--AUC
0.538; including them in the secondary comparison raises ROC--AUC to 0.562. A descriptive
ten-term sparse LDA axis separates the observed PHEME true- and false-labeled
groups with Cohen's $d=0.832$, and scientific retrieval can place external
evidence beside these early signals.

Together, these components form the proposed triage system and support a clear division of labor. Models forecast
subsequent recorded growth, summarize social-response patterns, and retrieve relevant
sources; human reviewers evaluate the claim, evidence, and context. Future
work should validate the reply-state annotations against human judgments,
evaluate the frozen sparse axis prospectively on unseen events, and test the
framework across additional platforms.

\bibliographystyle{plainnat}
\bibliography{references}

@article{shannon1948,
  author  = {Claude E. Shannon},
  title   = {A Mathematical Theory of Communication},
  journal = {Bell System Technical Journal},
  volume  = {27},
  pages   = {379--423, 623--656},
  year    = {1948}
}

@article{triage_fibvid_2021,
author  = {Jisu Kim and Jihwan Aum and SangEun Lee and Yeonju Jang and Eunil Park and Daejin Choi},
title   = {{FibVID}: Comprehensive Fake News Diffusion Dataset during the {COVID-19} Period},
journal = {Telematics and Informatics},
volume  = {64},
pages   = {101688},
year    = {2021},
doi     = {10.1016/j.tele.2021.101688},
url     = {https://doi.org/10.1016/j.tele.2021.101688}
}

@article{triage_qwen25_2024,
author  = {An Yang and Baosong Yang and Beichen Zhang and Binyuan Hui and Bo Zheng and Bowen Yu and Chengyuan Li and Dayiheng Liu and Fei Huang and Haoran Wei and Huan Lin and Jian Yang and Jianhong Tu and Jianwei Zhang and Jianxin Yang and Jiaxi Yang and Jingren Zhou and Junyang Lin and Kai Dang and Keming Lu and Keqin Bao and Kexin Yang and Le Yu and Mei Li and Mingfeng Xue and Pei Zhang and Qin Zhu and Rui Men and Runji Lin and Tianhao Li and Tianyi Tang and Tingyu Xia and Xingzhang Ren and Xuancheng Ren and Yang Fan and Yang Su and Yichang Zhang and Yu Wan and Yuqiong Liu and Zeyu Cui and Zhenru Zhang and Zihan Qiu},
title   = {{Qwen2.5} Technical Report},
journal = {arXiv preprint arXiv:2412.15115},
year    = {2024},
eprint  = {2412.15115},
archivePrefix = {arXiv},
primaryClass  = {cs.CL},
doi     = {10.48550/arXiv.2412.15115},
url     = {https://arxiv.org/abs/2412.15115}
}

@article{fisher1936,
author  = {R. A. Fisher},
title   = {The Use of Multiple Measurements in Taxonomic Problems},
journal = {Annals of Eugenics},
volume  = {7},
number  = {2},
pages   = {179--188},
year    = {1936},
doi     = {10.1111/j.1469-1809.1936.tb02137.x}
}

@book{cohen1988,
author    = {Jacob Cohen},
title     = {Statistical Power Analysis for the Behavioral Sciences},
edition   = {Second},
publisher = {Lawrence Erlbaum Associates},
address   = {Hillsdale, NJ},
year      = {1988}
}

@article{li2016structural,
author  = {Angsheng Li and Yicheng Pan},
title   = {Structural Information and Dynamical Complexity of Networks},
journal = {IEEE Transactions on Information Theory},
volume  = {62},
number  = {6},
pages   = {3290--3339},
year    = {2016},
doi     = {10.1109/TIT.2016.2555904}
}

@article{goel2016structural,
author  = {Sharad Goel and Ashton Anderson and Jake Hofman and Duncan J. Watts},
title   = {The Structural Virality of Online Diffusion},
journal = {Management Science},
volume  = {62},
number  = {1},
pages   = {180--196},
year    = {2016},
doi     = {10.1287/mnsc.2015.2158}
}

@inproceedings{zhao2015seismic,
author    = {Qingyuan Zhao and Murat A. Erdogdu and Hera Y. He and Anand Rajaraman and Jure Leskovec},
title     = {{SEISMIC}: A Self-Exciting Point Process Model for Predicting Tweet Popularity},
booktitle = {Proceedings of the 21th ACM SIGKDD International Conference on Knowledge Discovery and Data Mining},
pages     = {1513--1522},
publisher = {Association for Computing Machinery},
address   = {Sydney, NSW, Australia},
year      = {2015},
doi       = {10.1145/2783258.2783401}
}

@article{zubiaga2016analysing,
author  = {Arkaitz Zubiaga and Maria Liakata and Rob Procter and Geraldine Wong Sak Hoi and Peter Tolmie},
title   = {Analysing How People Orient to and Spread Rumours in Social Media by Looking at Conversational Threads},
journal = {PLOS ONE},
volume  = {11},
number  = {3},
pages   = {e0150989},
year    = {2016},
doi     = {10.1371/journal.pone.0150989},
url     = {https://doi.org/10.1371/journal.pone.0150989}
}

@inproceedings{kochkina2018all,
author    = {Elena Kochkina and Maria Liakata and Arkaitz Zubiaga},
title     = {All-in-one: Multi-task Learning for Rumour Verification},
booktitle = {Proceedings of the 27th International Conference on Computational Linguistics},
pages     = {3402--3413},
publisher = {Association for Computational Linguistics},
address   = {Santa Fe, New Mexico, USA},
year      = {2018},
url       = {https://aclanthology.org/C18-1288/}
}

@inproceedings{wei2019modeling,
author    = {Penghui Wei and Nan Xu and Wenji Mao},
title     = {Modeling Conversation Structure and Temporal Dynamics for Jointly Predicting Rumor Stance and Veracity},
booktitle = {Proceedings of the 2019 Conference on Empirical Methods in Natural Language Processing and the 9th International Joint Conference on Natural Language Processing (EMNLP-IJCNLP)},
pages     = {4787--4798},
publisher = {Association for Computational Linguistics},
address   = {Hong Kong, China},
year      = {2019},
doi       = {10.18653/v1/D19-1485},
url       = {https://aclanthology.org/D19-1485/}
}

@article{prollochs2021emotion,
author  = {Nicolas Pröllochs and Dominik Bär and Stefan Feuerriegel},
title   = {Emotions Explain Differences in the Diffusion of True vs. False Social Media Rumors},
journal = {Scientific Reports},
volume  = {11},
number  = {1},
pages   = {22721},
year    = {2021},
doi     = {10.1038/s41598-021-01813-2},
url     = {https://doi.org/10.1038/s41598-021-01813-2}
}

@inproceedings{peng2025rumor,
author    = {Xingyu Peng and Junran Wu and Ruomei Liu and Ke Xu},
title     = {Rumor Detection on Social Media with Temporal Propagation Structure Optimization},
booktitle = {Proceedings of the 31st International Conference on Computational Linguistics},
pages     = {3865--3878},
publisher = {Association for Computational Linguistics},
address   = {Abu Dhabi, UAE},
year      = {2025},
url       = {https://aclanthology.org/2025.coling-main.261/}
}

@inproceedings{chen2016xgboost,
author    = {Tianqi Chen and Carlos Guestrin},
title     = {{XGBoost}: A Scalable Tree Boosting System},
booktitle = {Proceedings of the 22nd ACM SIGKDD International Conference on Knowledge Discovery and Data Mining},
pages     = {785--794},
publisher = {Association for Computing Machinery},
address   = {San Francisco, CA, USA},
year      = {2016},
doi       = {10.1145/2939672.2939785}
}

@inproceedings{hassan2017claimbuster,
author    = {Naeemul Hassan and Fatma Arslan and Chengkai Li and Mark Tremayne},
title     = {Toward Automated Fact-Checking: Detecting Check-worthy Factual Claims by {ClaimBuster}},
booktitle = {Proceedings of the 23rd ACM SIGKDD International Conference on Knowledge Discovery and Data Mining},
pages     = {1803--1812},
publisher = {Association for Computing Machinery},
address   = {Halifax, NS, Canada},
year      = {2017},
doi       = {10.1145/3097983.3098131}
}

@inproceedings{mendes2023human,
author    = {Ethan Mendes and Yang Chen and Wei Xu and Alan Ritter},
title     = {Human-in-the-loop Evaluation for Early Misinformation Detection: A Case Study of {COVID}-19 Treatments},
booktitle = {Proceedings of the 61st Annual Meeting of the Association for Computational Linguistics (Volume 1: Long Papers)},
pages     = {15817--15835},
publisher = {Association for Computational Linguistics},
address   = {Toronto, Canada},
year      = {2023},
doi       = {10.18653/v1/2023.acl-long.881},
url       = {https://aclanthology.org/2023.acl-long.881/}
}

@inproceedings{wang2023checkcovid,
author    = {Gengyu Wang and Kate Harwood and Lawrence Chillrud and Amith Ananthram and Melanie Subbiah and Kathleen McKeown},
title     = {Check-{COVID}: Fact-Checking {COVID}-19 News Claims with Scientific Evidence},
booktitle = {Findings of the Association for Computational Linguistics: ACL 2023},
pages     = {14114--14127},
publisher = {Association for Computational Linguistics},
address   = {Toronto, Canada},
year      = {2023},
doi       = {10.18653/v1/2023.findings-acl.888},
url       = {https://aclanthology.org/2023.findings-acl.888/}
}

@inproceedings{lewis2020rag,
author    = {Patrick Lewis and Ethan Perez and Aleksandra Piktus and Fabio Petroni and Vladimir Karpukhin and Naman Goyal and Heinrich Küttler and Mike Lewis and Wen{-}tau Yih and Tim Rocktäschel and Sebastian Riedel and Douwe Kiela},
title     = {Retrieval-Augmented Generation for Knowledge-Intensive {NLP} Tasks},
booktitle = {Advances in Neural Information Processing Systems},
volume    = {33},
pages     = {9459--9474},
publisher = {Curran Associates, Inc.},
year      = {2020},
url       = {https://proceedings.neurips.cc/paper/2020/hash/6b493230-Abstract.html}
}

@inproceedings{cormack2009rrf,
author    = {Gordon V. Cormack and Charles L. A. Clarke and Stefan Büttcher},
title     = {Reciprocal Rank Fusion Outperforms Condorcet and Individual Rank Learning Methods},
booktitle = {Proceedings of the 32nd International ACM SIGIR Conference on Research and Development in Information Retrieval},
pages     = {758--759},
publisher = {Association for Computing Machinery},
address   = {Boston, MA, USA},
year      = {2009},
doi       = {10.1145/1571941.1572114}
}

@inproceedings{reimers2019sbert,
author    = {Nils Reimers and Iryna Gurevych},
title     = {Sentence-{BERT}: Sentence Embeddings using {S}iamese {BERT}-Networks},
booktitle = {Proceedings of the 2019 Conference on Empirical Methods in Natural Language Processing and the 9th International Joint Conference on Natural Language Processing (EMNLP-IJCNLP)},
pages     = {3982--3992},
publisher = {Association for Computational Linguistics},
address   = {Hong Kong, China},
year      = {2019},
doi       = {10.18653/v1/D19-1410},
url       = {https://aclanthology.org/D19-1410/}
}

\appendix

\section{Additional data details}
\label{app:data}

\begin{table}[ht]
\centering
\caption{Event and veracity composition of the 563-thread PHEME sample.}
\small
\begin{tabular}{@{}lrrr@{}}
\toprule
Event & False & True & Total \\
\midrule
Sydney Siege & 50 & 179 & 229 \\
Ottawa Shooting & 21 & 131 & 152 \\
Charlie Hebdo & 46 & 75 & 121 \\
Germanwings crash & 29 & 22 & 51 \\
Ferguson & 3 & 6 & 9 \\
Gurlitt & 0 & 1 & 1 \\
\midrule
Total & 149 & 414 & 563 \\
\bottomrule
\end{tabular}
\end{table}

\section{Factual-alignment feature comparison}
\label{app:factual-ablation}

Table~\ref{tab:factual-ablation} compares the primary event-held-out screen,
which excludes the explicit factual-alignment dimension, with the matched
secondary run that includes it. The cohort, folds, preprocessing,
regularization, and decision threshold are identical.

\begin{table}[ht]
\centering
\caption{Event-held-out comparison without and with factual-alignment-derived features. Candidate counts are reported before within-fold variance filtering and sparse regularization.}
\label{tab:factual-ablation}
\small
\setlength{\tabcolsep}{3.5pt}
\begin{tabularx}{\textwidth}{@{}Yrrrrr@{}}
\toprule
Feature family & Candidates & ROC--AUC & PR--AUC & Bal.\ acc. & Macro-F1 \\
\midrule
State composition, without & 94 & 0.538 & 0.764 & 0.522 & 0.484 \\
State composition, with & 104 & 0.562 & 0.774 & 0.539 & 0.501 \\
All response and lexical, without & 1,209 & 0.488 & 0.742 & 0.467 & 0.463 \\
All response and lexical, with & 1,305 & 0.543 & 0.765 & 0.545 & 0.532 \\
Combined, without & 1,217 & 0.492 & 0.745 & 0.471 & 0.465 \\
Combined, with & 1,313 & 0.546 & 0.767 & 0.549 & 0.536 \\
\bottomrule
\end{tabularx}
\end{table}

\clearpage
\section{Virality feature definitions and test predictions}
\label{app:virality}

\begin{table}[H]
\centering
\caption{The three predictors used in the FibVID propagation-tree continuation ablation. All are calculated at 30 minutes from the tree root.}
\small
\begin{tabularx}{\textwidth}{@{}lY@{}}
\toprule
Feature & Definition at the 30-minute observation cutoff \\
\midrule
Observed node count & number of recorded nodes in the snapshot, including the root \\
Structural entropy & unnormalized Shannon entropy in bits over observed node depths, including the root; zero for a root-only snapshot \\
Temporal entropy & unnormalized Shannon entropy in bits over positive-time arrivals in six fixed five-minute bins; zero when no such arrival is observed \\
\bottomrule
\end{tabularx}
\end{table}

\begin{figure}[H]
\centering
\includegraphics[width=0.98\textwidth]{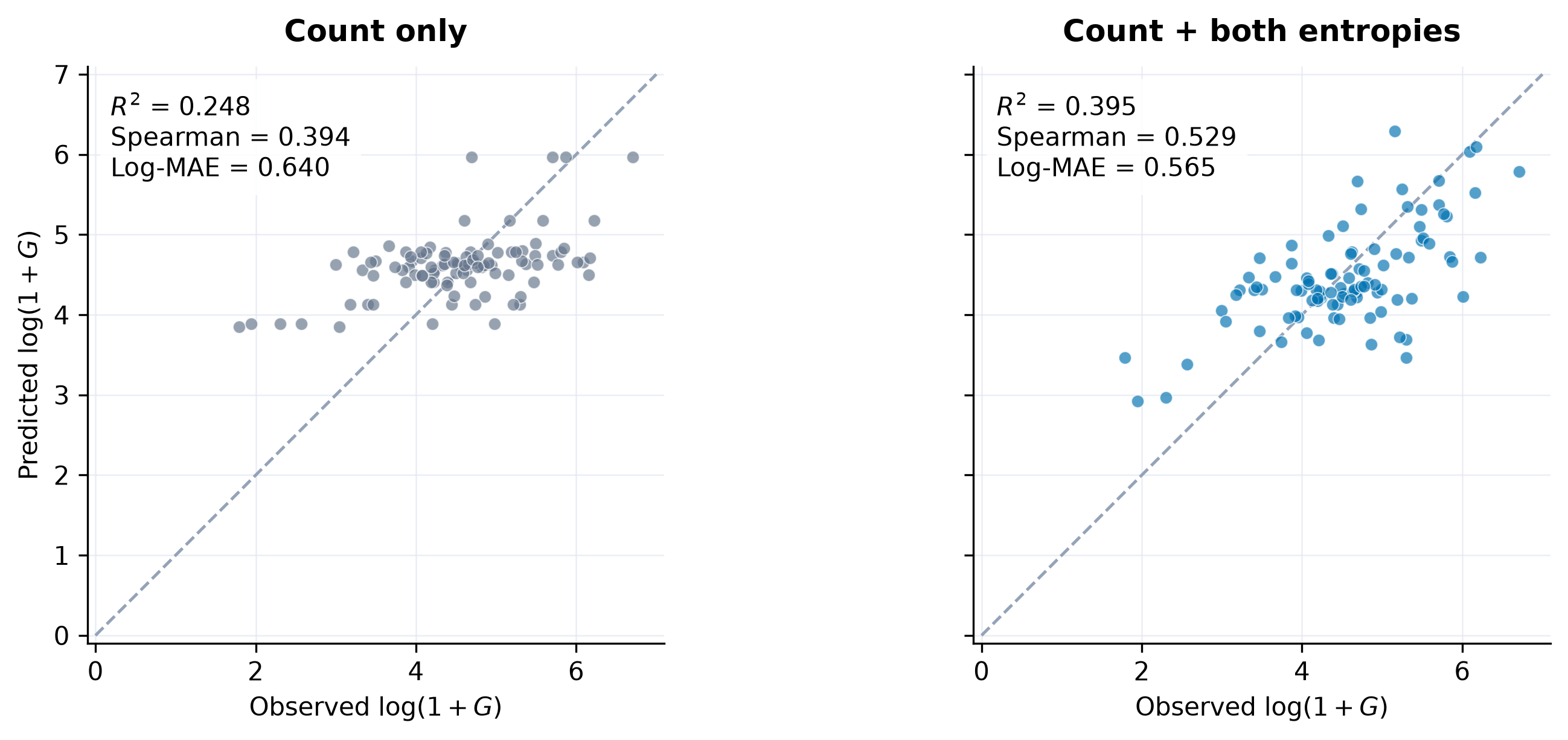}
\caption{Predicted versus observed $\log(1+G_c)$ for the 97 high-activity FibVID test propagation trees. Left: count-only baseline. Right: count plus structural and temporal entropy. Both models are fitted on 1,422 trees from 236 other claims. The dashed line denotes perfect prediction.}
\end{figure}

\clearpage

\section{Sparse-axis diagnostic classification}

Although the sparse LDA is used descriptively, we report its full-sample
threshold behavior for transparency. At the default 0.5 posterior-probability
threshold, the refitted model correctly assigns 31 of 149 false threads and
394 of 414 true threads. It assigns 118 false threads to the true side and 20
true threads to the false side. The corresponding in-sample ROC--AUC is 0.720,
PR--AUC is 0.872, macro-F1 is 0.580, and balanced accuracy is 0.580. These are
resubstitution diagnostics from the same cohort used for feature selection
and fitting; they must not be interpreted as held-out classification
performance. The empirical class prior and default threshold strongly favor
the majority true class.
\section{Retrieval curves}

\begin{figure}[ht]
\centering
\includegraphics[width=0.98\textwidth]{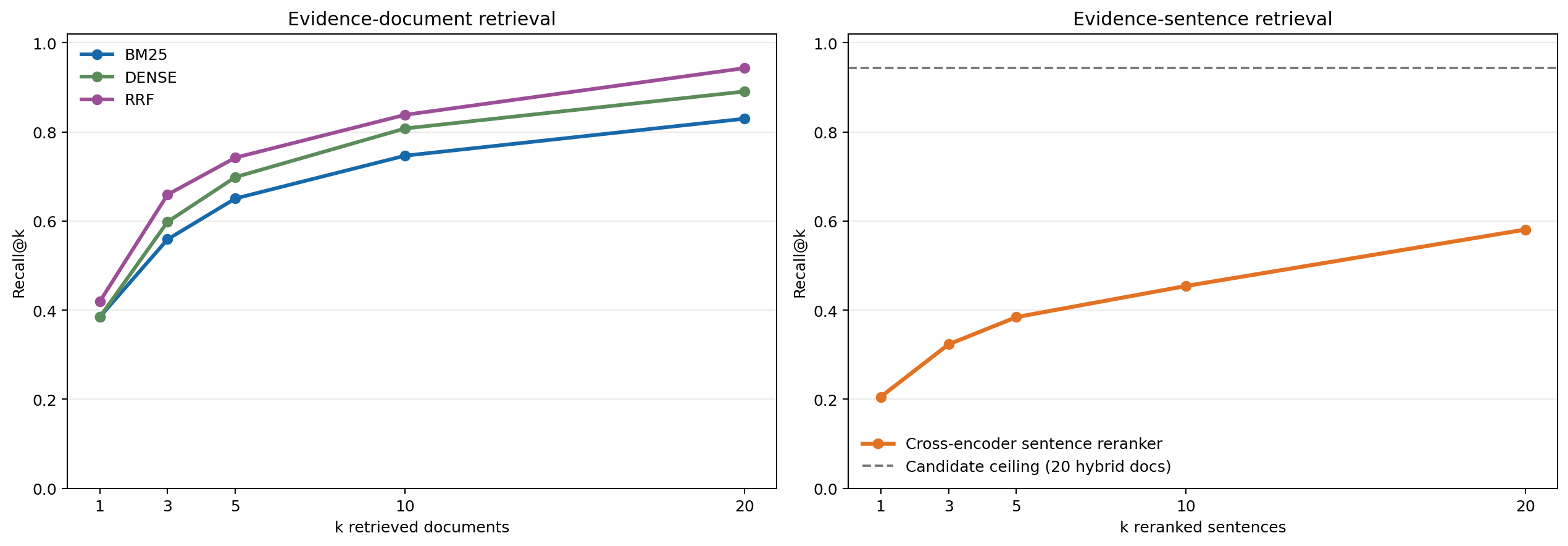}
\caption{Check-COVID evidence-document retrieval (left) and evidence-sentence reranking (right). The dashed line is the maximum sentence recall permitted by the 20-document candidate set.}
\end{figure}

\clearpage
\section{Qwen reply-state annotation prompt}
\label{app:qwen-prompt}

The following system message and user template are reproduced from the final annotation implementation in \texttt{cascaded\_pipeline\_aws\_v2.ipynb}. The prompt template is versioned \texttt{factorized-emotion-v2}; the \texttt{factorized-emotion-v3} identifier in the saved annotation metadata denotes a subsequent generation run rather than a different prompt. The placeholders are filled with the root claim, immediate parent message, and reply text. For a direct reply to the root, the template supplies the fallback shown below when no parent text is provided. The messages are passed through the tokenizer's chat template with an assistant-generation prefix.

\medskip
\noindent\textbf{System message.}\par
\begin{lstlisting}[basicstyle=\small\ttfamily,columns=fullflexible,breaklines=true,breakatwhitespace=true,keepspaces=true,showstringspaces=false]
Output only strictly valid JSON matching the requested schema.
\end{lstlisting}

\medskip
\noindent\textbf{User template.}\par
\begin{lstlisting}[basicstyle=\small\ttfamily,columns=fullflexible,breaklines=true,breakatwhitespace=true,keepspaces=true,showstringspaces=false]
You are coding observable discourse behavior in a social-media reply thread. Do not infer hidden motives. Use only the text and context supplied.

ROOT CLAIM:
{source_text}

IMMEDIATE PARENT MESSAGE:
{parent_text if parent_text else '[The reply is directly to the root claim.]'}

REPLY TO LABEL:
{reply_text}

Return exactly one valid JSON object with these keys and one allowed value per key:

1. root_stance: ENDORSE | REJECT | SUSPEND | MIXED | NO_STANCE
   Position toward the ROOT CLAIM.

2. parent_stance: ENDORSE | REJECT | SUSPEND | MIXED | NO_STANCE
   Position toward the IMMEDIATE PARENT MESSAGE. If directly replying to the root, this can match root_stance.

3. communicative_act: ASSERT | ARGUE | VERIFY | SOURCE_EVAL | RECONSTRUCT | TESTIFY | SOCIAL_META
   ASSERT = bare assertion, affirmation, or denial.
   ARGUE = gives reasons, evidence interpretation, logic, or methodological criticism.
   VERIFY = asks for or provides a checkable source, link, clarification, or confirmation.
   SOURCE_EVAL = evaluates competence, trustworthiness, authority, bias, or motive of a source.
   RECONSTRUCT = supplies an alternative causal account or replacement explanation.
   TESTIFY = personal experience, eyewitness account, or anecdote as evidence.
   SOCIAL_META = emotion-first reaction, identity signal, ridicule, praise, moral enforcement, tagging, sharing/reporting, or thread management.

4. factual_alignment: ACCURATE | INACCURATE | MIXED_UNRESOLVED | NO_FACTUAL_CLAIM | CANNOT_DETERMINE
   Use CANNOT_DETERMINE whenever the supplied context is insufficient to verify a factual assertion. Do not pretend to fact-check from memory.

5. primary_emotion: NEUTRAL | ANGER | FEAR_ANXIETY | SADNESS | DISGUST | SURPRISE | JOY_HOPE | TRUST_REASSURANCE | CONTEMPT_RIDICULE | OUTRAGE_MORAL_CONDEMNATION | EMPATHY_COMPASSION | OTHER_MIXED
   Choose the dominant explicitly expressed or strongly linguistically signaled emotion.

6. emotion_valence: POSITIVE | NEGATIVE | MIXED | NEUTRAL
7. emotion_intensity: NONE | LOW | MEDIUM | HIGH
8. emotion_target: NONE | CLAIM_EVENT | SOURCE_AUTHOR | PERSON_GROUP | INSTITUTION | SELF_INGROUP | OTHER
9. affective_claim: NONE | FEELING_EXPRESSION | THREAT_HARM | BLAME_ACCUSATION | MORAL_JUDGMENT | HOPE_REASSURANCE | EMPATHY_SOLIDARITY
   This is the main emotional/social proposition. Use NONE if the reply is not making an affective claim.

Return JSON only. No markdown, explanation, or extra keys.
\end{lstlisting}

\end{document}